# Spray-Coated Graphene/Quantum Dots Paper-Based Photodetectors

S. Malik[1], Y. Zhao[1], Y. He[1], X. Zhao[1], H. Li[1], L. G. Occhipinti[2], M. Wang[1], and S. Akhavan[1*]

[1]Institute for Materials Discovery, University College London, Torrington Place, London, WC1E 7JE, UK

[2]Cambridge Graphene Centre, University of Cambridge, JJ Thompson Avenue, Cambridge CB3 0FA, UK

## Abstract

Paper is an ideal substrate for the development of flexible and environmentally sustainable ubiquitous electronic systems. When combined with nanomaterials, it can be harnessed for various Internet-of-Things applications, ranging from wearable electronics to smart packaging. In this study, we present a non-vacuum spray deposition of arrays of hybrid single layer graphene (SLG)-$CsPbBr_3$ perovskite quantum dots (QDs) photodetectors on a paper substrate. This approach combines the advantages of two large-area techniques: chemical vapor deposition (CVD) and spray-coating. The first technique allows for the pre-deposition of CVD SLG, while the second enables the spray coating of a mask to pattern CVD SLG, electrode contacts, and photoactive QDs layers. The prepared paper-based photodetectors achieved an external responsivity of 520 A/W under 405 nm illumination at <1V operating voltage. By fabricating arrays of photodetectors on a paper substrate in the air, this work highlights the potential of this scalable approach for enabling ubiquitous electronics on paper.

## Introduction

The utilization of paper as a substrate presents a range of essential characteristics, encompassing sustainability, environmental friendliness, ubiquity, low cost, flexibility, biodegradability, recyclability, and deformability [1]. Bendable substrates are of great attraction in modern society due to their integration with non-flat surfaces that can allow for new device architectures. Traditional PD substrates such as silicon wafers, typically demand intense fabrication techniques and associated high energy costs. Thus, the use of paper is attractive, offering affordability for mass production alongside a scalable approach for commercialisation [2]. This is further reflected through the ease of fabrication, where unlike traditional electronics that require complex equipment and elevated temperatures, paper offers compatibility among a variety of printing techniques and inks for simple architectures [3], [4]. Furthermore, the paper-based substrate exhibits exceptional thermal stability in comparison to certain flexible plastic foils. The paper provides an excellent platform for the integration of two-dimensional (2d) and layered materials [5]. Paper is utilized as a versatile medium in various Internet-of-Things applications [6], including wearable electronics [7], smart packaging [8], and sensor-based systems such as humidity sensors [9] and pressure sensors [10]. However, paper is still a challenging substrate for electronics and has been rarely used without the inclusion of additional coating or laminating layers. The rough and porous nature, low stability, and endurance (primarily due to insufficient resistance to heat and humidity), in addition to the absence of established fabrication methods, are impeding its utilization in wearable applications [11], [12].

The printing of patterns on the paper substrate can be accomplished through various techniques, including screen printing [13], gravure printing [14], flexography [15], and inkjet printing [16]. Among these methods, inkjet printing stands out as one of the most promising due to its advantageous attributes, such as direct patterning without the need for masks [16] and high-resolution capabilities [17]. Nevertheless, the development of printed electronics, has primarily been constrained by the performance characteristics of the materials utilized as the constituents of these devices. Adhesion of the 2d inks to flexible substrates is a major challenge in device fabrication. The use of polymeric binders is restricted since binders limit 2d material ink functionalities [18], [19].

S. Conti et al. demonstrated the development of a field-effect transistor on a paper substrate through the integration of inkjet printing and chemical vapor deposition (CVD) [20]. This innovative approach employed CVD deposited layered materials to address the limitations associated with solution-processed materials, specifically those derived from liquid-phase exfoliation techniques of 2d materials [21], [22]. The fabrication process involved an initial step of optical lithography for patterning the CVD layer, followed by wet transferring and inkjet printing [20]. However, the reliance on complex and expensive fabrication tools, such as optical lithography, underscores the need for more accessible and cost-effective techniques to promote the commercialization and sustainability of paper-based electronics. In a recent study, Ref. [23] demonstrated inkjet lithography for the fabrication of photodetectors (PDs) utilizing CVD single-layer graphene (SLG) and black phosphorus ink on a $Si/SiO_2$ substrate. This approach resulted in the creation of PDs exhibiting external responsivity of 337 A/W at 488 nm. Notably, all fabrication processes were exclusively conducted via inkjet printing technology [23]. However, it is essential to acknowledge that the limitations of this technique include the relatively slower inkjet print speed and the associated higher operational costs, particularly concerning issues such as nozzle clogging that may occur with certain ink formulations [24], [19].

Spray coating is a mature and cost-effective technology offers numerous advantages over different coating techniques. It provides uniform coverage, fast application, and versatility for various substrates, including complex shapes and irregular surfaces, reducing material waste, and enabling precise control of layer thickness [25], [26]. Overall, spray coating stands out for its efficiency, adaptability, and consistent and controlled coating [25], [27]. Unlike spin-coating or doctor-blading methods, spray coating is not limited by the shape of the substrate. Additionally, the precursor solution is atomized into microsize droplets during spraying, so the deposition of the subsequent layer does not dissolve the previous ones [28]. This allows for fabrication of heterostructures made of different layers, and uniform film thicknesses of hundreds of micrometers or even millimeters to be achieved [29]

Paper-based PDs are lightweight, flexible, and cost-effective in nature, making them accessible for diverse applications. Paper-based PDs exhibit limitations in terms of responsivities, response time, and operating voltage due to several factors. The inherent properties of paper, such as its rough surface and light-scattering nature, can affect the efficiency of light absorption and conversion, resulting in lower responsivities compared to traditional PDs [30]. Table 1 demonstrates the current state-of-the-art paper-based PDs and their respective performance. Ref [31] presents an alternative all sprayed-processable Perovskite/MXene PD with responsivity of 0.049 A/W under a 10V bias at 450 nm. The high operating voltage of paper-based PDs requires higher electrical fields for optimal device performance. Thus, benefits of a lower power consumption can be observed in situations where continuous monitoring is required such as environmental sensors or long-term medical assessments. The highest responsivity achieved by a paper-based PD currently, uses $MAPbBr_3$ and can achieve a maximum responsivity of 1.3 A/W under a 1V bias at 365 nm [32].

These limitations pose challenges for achieving high-performance paper-based systems such as photodetection. Motivated by the benefits of quantum dots (QDs) in photodetection [33], [34], [35] and their integration with graphene to enhance the photoconductive gain mechanism [36], [37], this study introduces a novel class of paper-based PDs employing CVD SLG and QDs. Unlike other paper-based electronics that use additional, coating/laminating layers [38], [39], we have developed a technology based on scalable spray coating on a commercial paper substrate. Our results demonstrate high external responsivity of 520 A/W at 405 nm in the visible range, and low operating voltage (<1V), the highest reported (external responsivity) to date for paper-based PDs, to the best of our knowledge, see table 1. These findings highlight the potential of CVD layered-based materials for achieving high-performance photodetection in paper-based systems.

**Results and Discussion**

The fabrication process flow for QDs/SLG PD on paper substrate is outlined in Fig. 1. Canon A5 paper was chosen as the substrate to ensure the flexibility of the PDs. We carefully cut the entire sheet of paper into approximately 2 cm x 2 cm rectangular pieces. Afterward, we proceeded to clean these cut paper pieces by dripping Milli-Q water, followed by acetone and isopropanol, and nitrogen gas to get dried. A wet-transfer method was used to transfer the SLG from Cu foil to paper substrate. A polymethyl methacrylate (PMMA)/anisole (10 wt%) solution was spin coated at 3000rpm, 20s on top of the copper SLG, as a supporting layer. To etch the backside SLG on Cu foil, plasma etching was employed for 20 sec at 30 W. After etching the backside graphene, copper sheets were left to float on the ammonium persulfate solution in water with PMMA on the top. The resulting SLG/PMMA membrane were placed in DI water to clean the ammonium persulfate residuals and then transferred onto the paper substate, dried overnight and washed with acetone and IPA to remove PMMA.

To create the required pattern for spraying (Clarke, air brush kit), we designed a mesh screen with a side length of 0.81 mm as the frame of the mask, Fig. 1a. When spraying CNTs, the procedure must be conducted on a hot plate. The solvent used in the CNT solution is DMF, with a boiling point of 151 °C. To ensure rapid solvent evaporation after spraying, allowing efficient adhesion of CNTs to the SLG while avoiding excessively high temperatures that could impact sample performance, we set the substrate temperature to 100 °C. The spraying distance was maintained at 20 cm, and the spraying duration for 10 seconds to create an optimal electrode. Subsequently, the sample is dried at 100 °C for 20 min. Then, 5 wt% Polystyrene (PS) solution was dissolved in toluene solvent at 40 °C and spray-coated through the designed mask, Fig. 1b. Afterwards, The SLG was etched using plasma etching (40 sec at 30 W) to pattern the SLG while the SLG-based channel was protected by PS, Fig. 1c. The sample was then rinsed with water, followed by acetone and IPA, to eliminate the protective PS layer on top of the channel. Subsequently, QDs were applied via spray coating (Fig. 1d) with a target thickness of approximately 500 nm (Supplementary Information 1). The substrate temperature was set to 100 °C, and a short-burst spraying technique was employed to control the thickness. The process involved spraying 30 times over short, 2-3 seconds, durations to promote the formation of a consistent film. The sample dried rapidly after each spray coating. The schematic and image of our QDs/SLG PDs on paper substrate is shown in Fig. 1e.

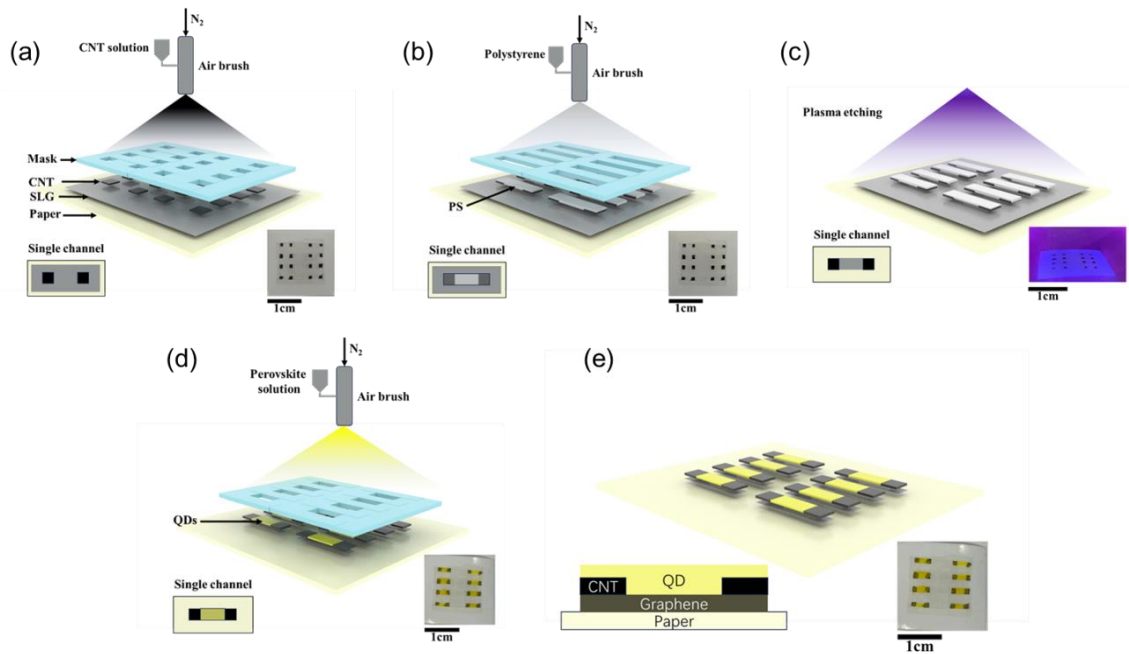

Fig. 1 The QDs/SLG PD fabrication steps on paper substrate. (a) CVD SLG is transferred on paper substrate, followed by overnight drying and PMMA removal using acetone/IPA. Then, CNT ink is spray-coated through the mask to make electrodes. The sample is placed on a hot plate at ~100 $^0$C for ~20 min. The image shows the spray-coated CNT ink after ~20 min annealing. (b) PS is spray-coated on SLG. The image shows spray-coated PS as mask on SLG. (c) SLG is then etched via Palma Etcher. The image shows the PS ink on SLG during plasma etching. (d) PS is removed by rinsing with water and then QDs are spray-coated through the mask. The image shows patterned SLG after removal of PS ink with water and spray-coated QDs. (e) Schematic and image of the QDs/SLG PDs on paper substrate.

The presence and quality of SLG are further studied by Raman spectroscopy. Raman spectra are acquired at 514.5nm using a Renishaw InVia with a 100X objective <0.1 mW. Fig. 2 plots the Raman spectrum of the film as grown on Cu after the Cu PL removal [53]. The 2D peak is a single Lorentzian with FWHM(2D)~40 cm$^1$, signature of SLG [54]. The position of the G peak, Pos(G), is~1591 cm$^{-1}$, with FWHM(G)~17 cm$^{-1}$. The 2D peak position, Pos(2D), is~2688 cm$^{-1}$, with FWHM(2D)~40 cm$^{-1}$, while the 2D to G peak intensity and area ratios, I(2D)/I(G) and A(2D)/A(G), are~3.4 and~7.9. Small D peak is observed at Pos(D)~1351 cm$^{-1}$, with I(D)/I(G)~ 0.1 indicating presence of small defects [55] of grown SLG on Cu foil.

In consideration of the significance of PS wettability on transferred SLG for patterning, the adhesion of PS was also tested. The PS ink was prepared by dissolving PS in toluene at 40⁰C to obtain a 5 wt% solution. The contact angle measurement revealed a value of 25.49⁰ and a surface tension of 235 mN/m for a drop of PS on SLG transferred onto the paper substrate (Supplementary Information 2). The wettability of our PS solution indicates its role as a protective layer that forms and good adherence to the SLG, safeguarding the underlying CVD SLG during plasma etching. The quality of SLG after spray-coated PS and etched CVD SLG is further investigated by Raman spectroscopy, Fig. 2. Pos(G)~1585 $cm^{-1}$, FWHM(G)~7 $cm^{-1}$, Pos(2D)~2691 $cm^{-1}$, FWHM(2D)~32 $cm^{-1}$, I(2D)/I(G)~0.8 A(2D)/A(G)~3.5, indicating p doping with $E_F$~ 120 meV by taking into account the dielectric constant~3 of the paper [56]. I(D)/I(G)~0.20 corresponds to a defect density~$5.8 \times 10^{10}$ $cm^{-2}$ at excitation energy of 2.41 eV and $E_F$~120 meV [57], [58].

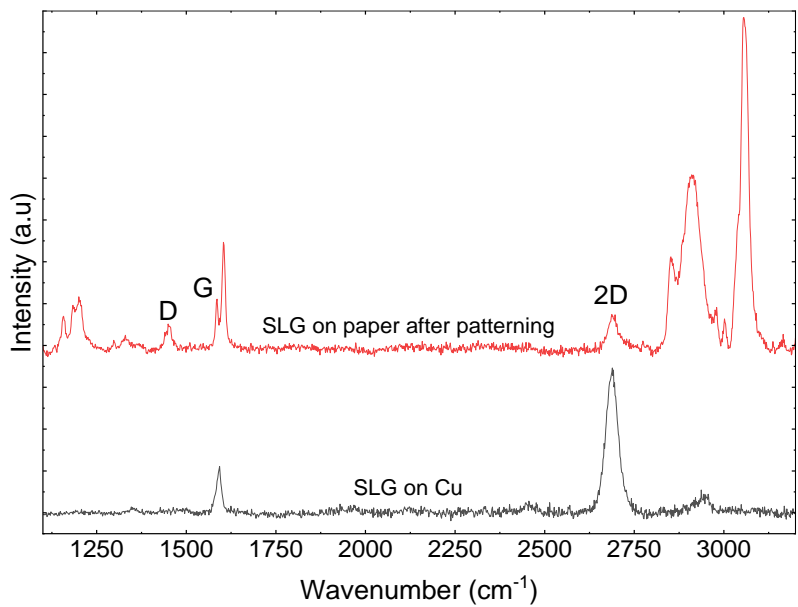

Fig 2. 514.5nm Raman spectra of CVD SLG on Cu and after patterning and PS removal on paper substrate.

The average sheet resistance of SLG on paper substrate, measured using a Four-point probe method, is 9.36 kΩ/sq indicating a conductivity of 173.19 kS/sq with our paper-based substrate coated with SLG layer. Ref [59] reported sheet resistance of transferred undoped SLG on a flexible PET substrate for SLG is 2.1 kΩ/sq [59]. The difference between our measured sheet resistance and the reported value could be attributed more to the roughness of our paper substrate than the PET counterpart.

Carbon nanotube (CNT) is dispersed in dimethylformamide (explained in detail in the methods section), and spray-coated as electrodes due to high conductivity and solution processable properties. In Fig. 3, the graph depicts the optical absorbance (Abs) of dispersed CNTs in dimethylformamide (DMF). The concentration of dispersed CNT is determined through the Beer-Lambert Law, which relates Abs to concentration (c (g/L)), the extinction coefficient ($\epsilon_{ext}$ (L/g.m)), and the cuvette length (L(m)) [48]. The solution was first diluted 150 times beforehand to prevent the saturation of the spectrometer's detector. In a previous study [49], the CNT $\epsilon_{ext}$ at 300 nm was experimentally derived by measuring the slope of Abs per length versus CNT concentration, resulting in an $\epsilon_{ext}$ value of approximately $5.625 \times 10^{-3}$ L/g.m. This estimation yielded a concentration of approximately c~0.63 g/L for our CNT ink, consistent with findings in prior research [49]. The presence and quality of CNT is further studied by Raman spectroscopy, Fig. 3b. Raman spectra are acquired at 514.5 nm using a Renishaw InVia with a 100× objective <0.5 mW. The three active modes for CNT are D, G, and 2D modes [50]. The G band is the in-plane bond stretching mode of the C-C bonds in the hexagonal lattice. The 2D peak is related to the Raman scattering due to a vibrational mode characterized by the breathing of six carbons pertaining to a hexagon in the hexagonal lattice. In the presence of a defect, the first order component of the hexagon-breathing mode is activated combined with an elastic scattering of a photo-excited electron by the defect, as a double-resonance Raman peak and it is called the D band [50]. The position of the G peak, Pos(G), is~1593 $cm^{-1}$, with FWHM(G)~76 $cm^{-1}$. The 2D peak position, Pos(2D), is~2701 $cm^{-1}$, with FWHM(2D) ~106 $cm^{-1}$, while the 2D to G peak intensity and area ratios, I(2D)/I(G) and A(2D)/A(G), are~0.32 and~0.46, respectively. D peak was observed at 1353 $cm^{-1}$ with I(D)/I(G)~1.02. Fourier-transform infrared spectroscopy (FTIR) was used to assess the functional groups in the CNTs. As shown in Fig. 3c, O-H bonds are observed near 3542 $cm^{-1}$, C-H bonds near 2933 $cm^{-1}$, C=O bonds near 1659 $cm^{-1}$, O-C=O bonds near 1386 $cm^{-1}$, and C-O bonds near 1092 $cm^{-1}$. In particular, the characteristic peak of Triton X-100 was observed near 651 $cm^{-1}$ (C=C) [51]. When dissolving CNTs with Triton X-100, there are some extra peaks or spectral changes that can occur since Triton X-100 has specific functional groups, such as alkenyl groups (C=C), which can produce additional characteristic peaks in the FTIR spectrum. At the same time, additional peaks of its same characteristic groups as CNT can superimpose with the spectrum of CNT, resulting in different characteristic peaks observed in the FTIR spectrum [52]. We measured the sheet resistance of the CNT ink coated on the paper substrate using a Four-Point probe setup, and the resultant sheet resistance was found to be 20.04 Ohm/sq. Scanning electron microscopy images were acquired to check the topography of the spray-coated CNT on paper substrate (Supplementary Information 3). The SEM image reveal the degree of dispersion and aggregation of coated CNTs on the paper substrate. To assess electrode feasibility, the wettability of CNT ink on transferred SLG over a paper substrate, is investigated. Optical contact angle and surface tension measurements (Attension) are employed for characterization. The contact angle measurement was found to be 29.51º and surface tension to be 224.3 mN/m for a drop of CNT on SLG transferred onto the paper substrate, allowing for swift drying with minimal spread outside the restricted area (Supplementary Information 2).

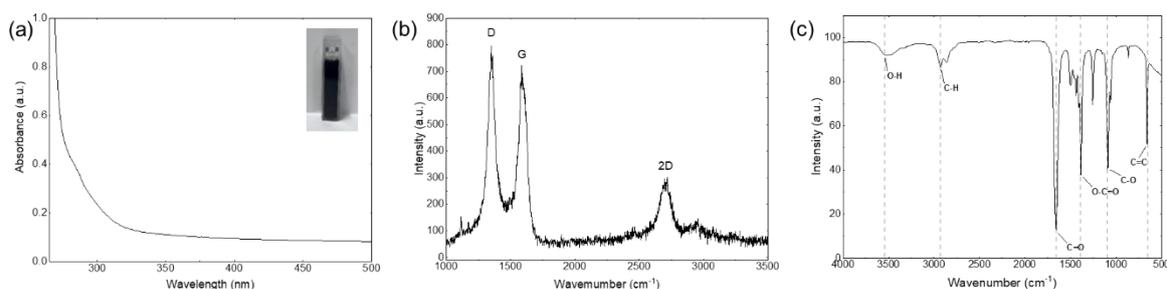

Fig. 3. (a) Absorbance of dispersed MWCNT in DMF. The dispersions are diluted to avoid detector saturation. Inset image shows dispersed CNT ink. (b) Raman spectra at 514.5nm of MWCNT. (c) The FTIR of MWCNT ink.

Stable $CsPbBr_3$ perovskite QDs were synthesized using a modified recipe from Ref. [40], which is explained in detail in the methods section. The synthesis of $CsPbBr_3$ perovskite QDs can be observed by a rapid and notable colour change reaction, which signifies its swift formation. Typically, within the first ten seconds of the reactants' contact, the reaction initiates, resulting in a gradual transition to a fluorescent yellow solution. The reaction reaches its completion in just two minutes, with no further product formation observed. The photoluminescence (PL) and optical absorption spectra were collected from purified $CsPbBr_3$ perovskite QDs. These measurements were performed on samples dispersed in toluene and on solid films deposited via spin-coating on glass, as depicted in Fig. 4a. The QDs in the solution exhibited a PL peak centred at approximately 516 nm, with a full-width-half-maximum (FWHM) like that of cubic 8.5 nm nanocrystals [41], [42]. In contrast, the thin film displayed a redshift (~4 nm) in the optical absorption edge, transitioning from 516 nm to 520 nm, as determined from the band edge. This shift may be attributed to several factors, such as surface defects, strain effects, and variations in size and shape [43], [44]. Notably, the PL observed in the film consistently aligned with the absorption edge, maintaining a consistent Stokes shift of approximately 6 nm and a FWHM like that observed in the solution. The Tauc plot (Fig. 4b) method was employed on the UV-Vis spectra to assess the bandgap of the QDs, resulting in values of 2.18 eV and 2.22 eV for QDs in solution and on glass, respectively.

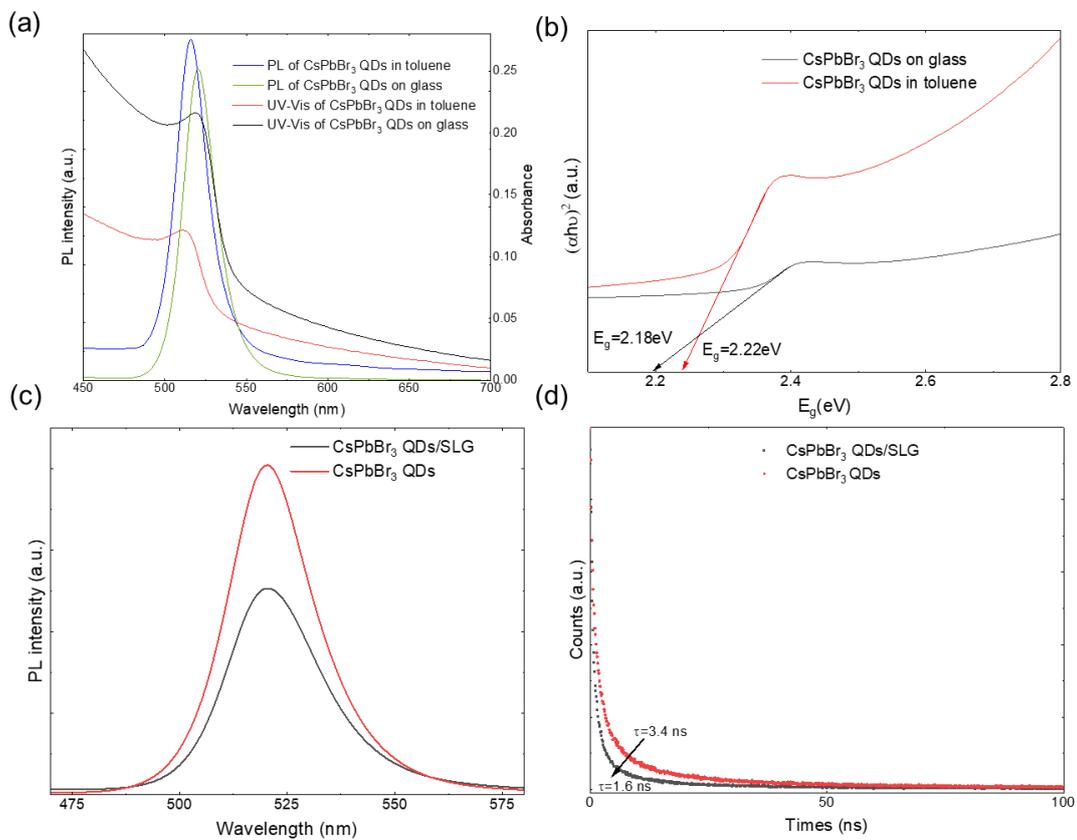

Fig. 4. (a) PL and UV-Vis spectra of $CsPbBr_3$ QDs coated on glass and dispersed in toluene. (b) Tauc-plot of the UV–vis absorption spectra and calculation of the bandgap energy of $CsPbBr_3$ QDs coated on glass and dispersed in toluene. (c) PL spectra of $CsPbBr_3$ QDs and $CsPbBr_3$ QDs/SLG on glass substrate. (d) TRPL spectra of $CsPbBr_3$ QDs and $CsPbBr_3$ QDs/SLG on glass substrate.

The PL spectra of QDs and QDs/SLG are shown in Fig. 4c. Both show a PL peak ~520 nm arising from the QDs band gap. The PL intensity (integrated area under PL curve) of QDs/SLG on glass is quenched ~33% compared to QDs on glass, Fig. 4c.

This can be assigned to charge carrier transfer between QDs and SLG [45]. Fig. 4d shows the time-resolved photoluminescence (TRPL) spectra of CsPbBr$_3$ QDs and CsPbBr$_3$ QDs/SLG both are on glass substrates to investigate the excited-state dynamics of the QDs/SLG. The CsPbBr$_3$ QDs on a glass substrate exhibited a biexponential decay pattern, consistent with prior literature [46], featuring an average fluorescence decay time of 3.4 ns. In contrast, the QDs/SLG structure displayed a notably shorter average fluorescence decay time of 1.6 ns, as depicted in Fig. 4d. Our findings indicate that SLG, in this context, exhibits quenching effects like those reported with perovskite QDs on SLG prepared via spin-coating techniques [47]. We believe that the PL quenching and lifetime shortening in TRPL for CsPbBr$_3$ QDs/SLG, point toward a predominant electron transfer mechanism from CsPbBr$_3$ QDs to SLG. The deposition of QDs on SLG appears to enhance charge transfer by facilitating π-π electron interactions between the QDs and the sp$^2$-hybridized SLG layer. In essence, the observed PL quenching suggests the occurrence of rapid charge transfer within the QDs/SLG superstructure [37]. Furthermore, a contact angle of 10.88° and a surface tension of 329.3 mN/m was measured for a drop of QDs on SLG (Supplementary Information 2). This small contact angle suggests a strong affinity of the SLG surface towards QDs. The wettability characteristics of QDs provide valuable insights into the formation of a compact film through their strong interaction with SLG.

The photoelectrical performance of the paper-based PDs was evaluated using Thorlabs laser diode, which emitted light at a wavelength of 405 nm. The laser beam, with a beam spot size of 5mm$^2$ in diameter, was employed as the excitation light and covered the entire SLG channel. The PD responsivity is a key parameter of PDs and can be defined either as external [60]:

$$R_{ext} = |I_{light} - I_{dark}|/(P_{opt}.A_{PD}/A_{opt})$$

or internal [60]:

$$R_{int} = |I_{light} - I_{dark}|/(P_{abs}.A_{PD}/A_{opt})$$

where $I_{light}$ and $I_{dark}$ are the currents of the PD under illumination and in dark conditions, respectively. $A_{PD}$ refers to the PD area, while $A_{opt}$ represents the laser spot size. The scaling factor $A_{PD}/A_{opt}$ considers that only a fraction of the optical power reaches photoactive area of PD. $P_{opt}$ denotes the incident optical power, and $P_{abs}$ refers to the absorbed optical power. In our measurements, the light spot had a diameter of 5 mm, corresponding to an area of 19.64 mm$^2$. The channel area was fabricated using a pattern size of 0.81 mm, yielding an area of 0.66 mm$^2$. It's worth highlighting that not all incoming photons are entirely absorbed by the PD, which causes the optical power output ($P_{opt}$) to surpass the absorbed power ($P_{abs}$). Consequently, the internal responsivity ($R_{int}$) is higher than the external responsivity ($R_{ext}$). Nonetheless, $R_{ext}$ offers a more comprehensive overview of PD performance as it accounts for various factors, including light reflection, reabsorption, material quality, etc [61].

Fig. 5a depicts the current in the channel (I) as a function of voltage in dark and under illumination at a wavelength of 405 nm with 0.51 mW optical power. We observed a linear relationship between the current through the device and the applied voltages, indicating a strong connection between the channel and electrodes. This linearity can be attributed to the Ohmic nature of the QDs/SLG/Ag junction. The Ohmic behaviour of the PD is advantageous as it facilitates the generation of high photocurrent. This behaviour allows charges to flow easily from the conduction band of SLG to the metal contacts, enhancing the device's performance. Upon illumination, we noticed an increase in the current, indicating hole (h) dominated carrier transport. This suggests the transfer of holes from QDs to SLG. The mechanism was elucidated based on band diagrams, which confirmed the p-type nature of SLG through Raman shift analysis, as showed in Fig. 2.

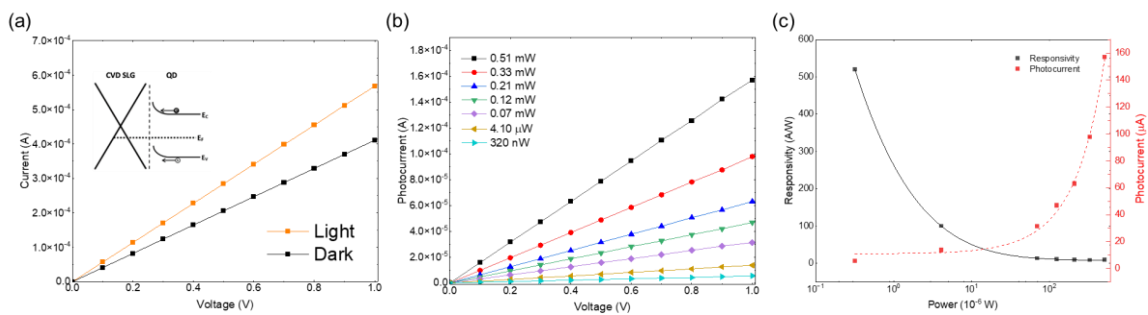

Fig. 5. (a) Current as a function of bias under dark and illumination, 405 nm and 0.51mW. The inset image, is the schematic band diagram of QDs/SLG interface, showing the QDs conduction band and valence band, generation of e/h pairs and transfer of h from QDs to SLG. (b) Photocurrent as a function of bias for different illumination powers. (c) $R_{ext}$ and photocurrent as a function of optical power density.

The movement of carriers described above, resulted in the generation of a local electric field at the SLG and QD interface, with the field direction pointing towards SLG. This electric field induced an upward band bending of QDs towards SLG, further influencing the behaviour of the system. Under light illumination, the top layer (QDs) absorbs the photons from incoming light, leading to the generation of electron-hole (e-h) pairs. Due to the presence of the local electric field, photogenerated electrons or holes transfer to SLG and subsequently move toward the electrodes. This phenomenon contributes to either an increase or decrease in the current, depending on the direction of charge flow. During illumination of our paper-based PD, light is absorbed by QDs, and photogenerated holes (h) are transferred from the valence band of the QDs to lower energy states in SLG. This transfer results in the accumulation of photogenerated electrons (e) as uncompensated charges in the system. The photogenerated electrons, being trapped in QDs, function as an additional negative gate when applied to the SLG channel. This alters the electric field at the junction between QDs and SLG. Fig. 5b plots the photocurrent as a function of $V_{ds}$. This is defined as:

$$I_{photo} = I_{light} - I_{dark}$$

where $I_{light}$ is the current under illumination and $I_{dark}$ is current in dark conditions. When $V_{ds}$ exceeds 1V, the drift velocity of free carriers ($v_d = \frac{\mu E}{1 + \mu E / v_{sat}}$) [60], where $v_{sat}$ represents the saturation velocity of carriers in the channel and E is the applied electric field to SLG, increases linearly until it reaches saturation, primarily due to carrier scattering with optical phonons. Consequently, all measurements are conducted with $V_{ds} \leq 1V$ to maintain the device's operation within the linear (Ohmic) regime, thus eliminating the nonlinear dependence of $v_d$ on $V_{ds}$, Fig. 5b. To derive $R_{ext}$, we measure $I_{photo}$ at different optical powers via an attenuator ranging from 510 µW to 0.32 µW, Fig. 5c. Fig. 5c illustrates an increase in $R_{ext}$, starting from 8.87 A/W, and reaching approximately 520 A/W at wavelength of 405 nm when $V_{ds}$ is at approximately 1V with an optical power of 0.32 µW. At an optical power of approximately 0.32 µW the number of photogenerated carriers decreases, resulting in an increase in the built-in field at the SLG and QDs interface. This increase explains the enhancement of $R_{ext}$ at lower optical powers. $I_{photo}$ increases with decreasing optical power, consistent with the findings reported by Refs [62], [63], [23]. This change in $R_{ext}$ can be explained by the shielding of the built-in electric field.

When light is illuminated into the photosensitive layer of the device, the generated electron-hole pairs are separated by the built-in electric field created at the interface between the SLG and QDs, causing electrons to transfer into the SLG and decrease its hole concentration, resulting in lower conductivity. At high light intensities, more electrons in SLG recombine with holes, resulting in a lower responsivity. However, as the light intensity decreases, fewer electrons are transferred into SLG, reflecting a higher responsivity.

Normalized Detectivity ($D^*$ ($cm \cdot Hz^{1/2}/W$ or Jones)) relates the performance of PDs in terms of $R_{ext}$ to the photoactive area of the PD, allowing the comparison of PDs with different active areas. The noise current in the shot noise limit is defined as $I_n = (2qI_{dark})^{1/2}$. Thus $D^*$ on the shot noise limit can be expressed as:

$$D^* = R_{ext}(AB)^{1/2}/(2qI_{dark})^{1/2}$$

Where B is the bandwidth and A is the photoactive area of the PD. $D^*$ is calculated as ~$3.602 \times 10^{12}$ Jones for our PD, which is 10 times greater than the highest reported detectivity of any current paper-based PD [64]. Ref [64] reports a self-biased PD that uses $Ga_2O_3$ and exhibits a detectivity of $1.42 \times 10^{11}$ Jones with a responsivity of $3.1 \times 10^{-3}$ A/W under a 254 nm illumination [64]. Ref. [32] reports a $R_{ext}$ of 1.3 A/W with a detectivity of $7.7 \times 10^{10}$ Jones under a 1V bias and 365 nm wavelength composed of $MAPbBr_3$. Thus, the device presented in this paper is superior through both detectivity and responsivity metrics, demonstrating great promise for this novel architecture and fabrication method. Table 1 shows different paper-based PDs from recent published work and their corresponding parameters: applied voltage, $R_{ext}$, operating wavelength, detectivity, and flexibility. Compared with other PDs, the responsivity of our work (520 A/W) at only a 1V bias shows great potential in flexible PD industry.

Table 1: A comparison of different paper-based PDs regarding operating voltage, external responsivity, operational wavelength, detectivity, and flexibility.

| Material used | Bias (V) | $R_{etx}$ (A/W) | Operational wavelength | Detectivity ($cm \cdot Hz^{1/2}/W$) | Bending cycles | Reference |
|---|---|---|---|---|---|---|
| **QDs/SLG** | **1** | **520** | **VIS** | **$3.6 \times 10^{12}$** | **500 +** | **This work** |
| Cu/Cu$_2$S | 0.1 | 0.027 | VIS NIR | $2.4 \times 10^8$ | 1000 | [65] |
| WSe$_2$ | 15 | $5.43 \times 10^{-3}$ | UV | - | 1000 | [66] |
| MoS$_2$ | 20 | $1 \times 10^{-7}$ | NIR | - | - | [67] |
| MoS$_2$ | 4 | $2.8 \times 10^{-4}$ | NIR | $1.8 \times 10^9$ | - | [68] |
| MoS$_2$/ZnS | 1 | $1.7 \times 10^{-6}$ | UV | - | 500 | [69] |
| MoS$_2$/ZnS | 1 | $9 \times 10^{-7}$ | VIS | - | 500 | [69] |
| MoS$_2$/ZnS | 1 | $4 \times 10^{-7}$ | NIR | - | 500 | [69] |
| ZnO | 10 | $12 \times 10^{-6}$ | UV | $1.23 \times 10^9$ | - | [70] |
| ZnO | 9 | 0.06 | UV | - | - | [71] |
| MAPbI$_3$ | 5 | $4.4 \times 10^{-3}$ | VIS | - | - | [72] |
| WS$_2$ | 10 | 0.439 | VIS | $1.41 \times 10^{10}$ | - | [73] |
| PVK-MXene | 10 | $4.49 \times 10^{-2}$ | VIS | $2.55 \times 10^9$ | - | [31] |
| WSe$_2$ | 5 | $1.778 \times 10^{-2}$ | VIS | $5.86 \times 10^{10}$ | - | [74] |
| MAPbBr$_3$ | 1 | 1.3 | UV | $7.7 \times 10^{10}$ | - | [32] |
| Ga$_2$O$_3$ | - | $3.1 \times 10^{-3}$ | UV | $1.42 \times 10^{11}$ | - | [64] |
| Au-NPs/MoS$_2$ | 5 | $9.93 \times 10^{-2}$ | VIS | $3.53 \times 10^{10}$ | 500 | [75] |
| Au-NPs/MoS$_2$ | 5 | $4.604 \times 10^{-2}$ | NIR | $2.17 \times 10^{10}$ | 500 | [75] |
| MoS$_2$/WSe$_2$ | 5 | 0.124 | VIS | - | 1000 | [76] |
| Bi$_2$S$_3$/RuS$_2$ | - | $5.28 \times 10^{-2}$ | VIS | $105 \times 10^9$ | - | [77] |
| Bi$_2$S$_3$/RuS$_2$ | - | $39 \times 10^{-2}$ | UV | $60 \times 10^9$ | - | [77] |
| Bi$_2$S$_3$/RuS$_2$ | - | $6 \times 10^{-3}$ | NIR | $15 \times 10^9$ | - | [77] |
| PbS | 1 | $6.45 \times 10^{-3}$ | NIR | $6.4 \times 10^{10}$ | 124 | [78] |
| ZnS/polyaniline | - | $3.671 \times 10^{-3}$ | UV | $3.1 \times 10^{10}$ | - | [79] |
| Ti$_3$C$_2$T$_x$-TiO$_2$ | 5 | $7.8 \times 10^{-5}$ | VIS/UV | $8.40 \times 10^4$ | - | [80] |
| PbS-QDs/SWCNTs | - | 0.07 | VIS | - | 1000 | [81] |
| WSe$_2$/PANI | 5 | $1.726 \times 10^{-2}$ | VIS | $1.11 \times 10^{10}$ | 400 | [82] |
| Te-NWs | - | $2.25 \times 10^{-2}$ | VIS | $3.67 \times 10^{10}$ | 500 | [83] |
| Te-NWs | - | $1.45 \times 10^{-2}$ | NIR | $2.37 \times 10^{10}$ | 500 | [83] |
| SnS$_2$/SLG | - | $6.89 \times 10^{-3}$ | VIS | $5.26 \times 10^9$ | 500 | [84] |
| SnS$_2$/SLG | - | $3.67 \times 10^{-3}$ | UV | $1 \times 10^{10}$ | 500 | [84] |

Flexibility is a characteristic of great desirability for PDs with wearable devices being a leading application for this technology. Therefore, the performance of our PD was tested over the course of a 500 bending cycle test to assess its effectiveness when subjected to cyclic strain. A bending radius of 4.5 mm was introduced, and a single point bending test was conducted. Both the dark current and current under 0.51 mW light were recorded with the device at rest (i.e. prior to bending), to calculate the photocurrent at rest ($I_{photo(Rest)}$). Following this, the device was then bent at a series of intervals up to 500 bending cycles. Fig. 6a plots $I_{photo(Bend)}/I_{photo(Rest)}$ as a function of bending cycles. It can be observed that the device demonstrated a decay in performance with increasing bends, such that after 500 bending cycles, $I_{photo(Bend)}/I_{photo(Rest)}$ drops~15% of initial cycle. This behaviour, where the device response initially experiences a downward drift for the first several cycles and subsequently maintains a stabilized trend, is also observed in other sensors [85], [86], [87]. This could be attributed to the construction of some new conductive networks and the subsequent formation of an equilibrium state. In our QDs/SLG PDs, it was worth mentioning across all bending cycles a photocurrent was observed and device retained its ohmic contact (linear photocurrent vs voltage), indicating the device remained functional throughout all tests, Fig. 6b. We believe improvements in this metric to come from an increase in the number of SLG layers used, as this would provide a more durable structure and allowing current to travel through other SLG layers should initial layers fail. This would result in a high constant photocurrent bend/photocurrent ratio and further justify the use case of this fabrication technique in the industry.

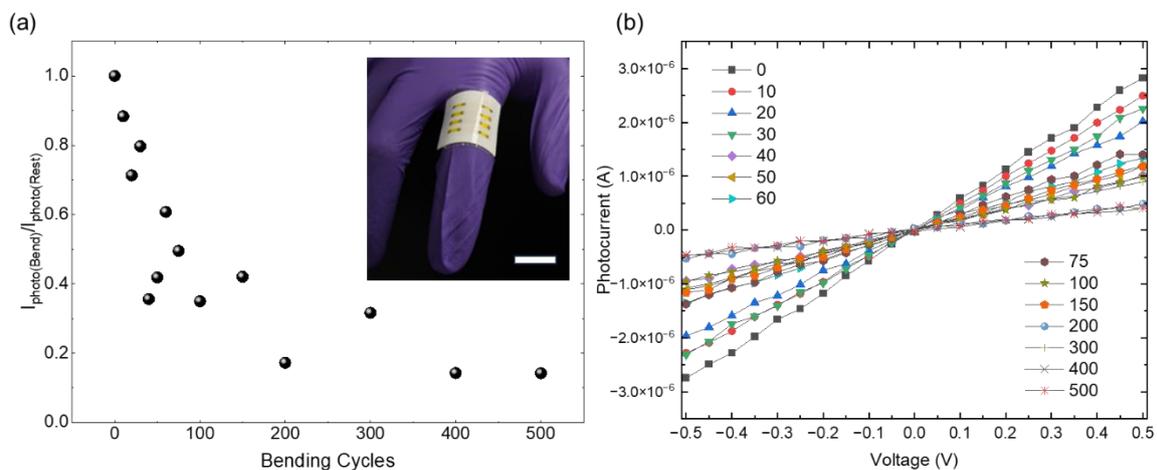

Fig. 6. (a) $I_{photo(Bend)}$ normalized to that measured on flat PD ($I_{photo(Rest)}$), as a function of bending radius. Inset image is the QDs/SLG PD array on paper substrate. Scale

bar is 1 cm. (b) $I_{photo(Bend)}$ normalized to that on flat PD ($I_{photo(Rest)}$) as a function of bending cycle.

## Conclusion

We have successfully demonstrated high-performance paper-based PDs composed of CVD SLG and $CsPbBr_3$ perovskite QDs via spray-coating technique under ambient conditions. We achieved a maximum external responsivity of 520 A/W at 405 nm with <1V operating voltage. Our device fabrication approach has proven suitable for the development of scalable electronics on a paper substrate. This work showcases the significant potential of non-vacuum fabrication of high-performance active devices for next-generation electronics on paper, enabling cost-efficient, environmentally friendly, and sustainable practical applications.

## Acknowledgement

We gratefully acknowledge the financial and technical support by University College London under the project 575095ZZN.

## Methods

Materials:

Lead (II) bromide ($PbBr_2$, 99.999% trace metals basis), caesium carbonate ($Cs_2CO_3$, ReagentPlus, 99%), butylamine (BuAm, 99.5%) 2-propanol (IPrOH, anhydrous, 99.5%), propionic acid (PrAc, ≥99.5%), n-hexane (HEX 99.5%), toluene (TOL, anhydrous, 99.8%), acetone (≥99.5%), isopropanol (70% in H2O), anisole (≥99%), and toluene (99.9%) were purchased from Sigma Aldrich. Polymethyl methacrylate (average Mw 97000), ammonium persulfate (≥98%), and polystyrene (average Mw ~280,000) were supplied from Sigma-Aldrich in powder form. MWCNTs were purchased from Sigma-Aldrich, Co., 3050 Spruce Street, St Louis, MO 63103USA with a purity of >90%, inner diameter of 110-170 nm, and length in the range of 5-9 μm. SLG grown on copper foil purchased from Graphenea. All chemicals were used without any further purification and Milli-Q water was used throughout.

Dispersion of MWCNT:

20 mg of MWCNTs and 10 mg of surfactant (Triton X-100) were added to 10 mL of a Dimethylformamide and then sonicated for 6 hours at 50°C. The process was followed by centrifugation of the suspension via Thermo Scientific Heraeus Multifuge X1 (2272 xg) at 4000 rpm for 30 min to remove nondispersible materials.

Synthesis of $CsPbBr_3$ perovskite QDs:

The synthesis procedure for $CsPbBr_3$ perovskite QDs followed a straightforward and rapid one-step injection method. It commences by dissolving $Cs_2CO_3$ in propionic acid (PrAc) to form a $Cs^+$ propionate complex. This complex is then diluted in a solvent mixture containing isopropanol (IPrOH), hexane (HEX), and butylamine (BuAm), comprising both polar and a polar component, all at room temperature. Notably, this dissolution reaction is exothermic, eliminating the need for external heating. Additionally, this method doesn't require degassing of the precursors. Simultaneously, another solution is prepared at ambient conditions, involving the dissolution of $PbBr_2$ in a similar chemical mixture (also at room temperature). This solution is subsequently injected into the first one. Within just 10 seconds after injection, the QDs initiate nucleation and rapidly reach their maximum size. Following this stage, the QDs are isolated through centrifugation and redispersed in toluene, making them ready for direct use in device fabrication.

Characterisation and Fabrication:

The Lambda 750S UV-Vis spectrometers (Perkin Elmer) were used to conduct measurements of optical absorption spectra at room temperature.
Fourier-transform infrared spectroscopy analysis to assess the functional groups was done by using Perkin Elmer Spectrum Two FT-IR Spectrometer. Oxford Instruments X-Max 80 (SDD) EDS system was utilized to check scanning electron microscopy images. Steady-state and time-resolved PL spectra were acquired on a Time-correlated single photon counting (TCSPC) (LifeSpec-ps) by Edinburgh Instruments, employing an excitation wavelength of 405 nm at -15 °C. Sample preparations for measurements were done through spin-coating on various substrates using the Ossila spin coater. SLG was etched using plasma etching via Diener electronic GmbH & Co. Raman spectra were gathered using the inVia Raman microscope (Renishaw) at room temperature, utilizing an excitation wavelength of 515 nm and a

100x objective lens. Performance tests for the photodetector were carried out using the B1500A Semiconductor Device Analyzer (Keysight). The sheet resistance of different layers of CVD graphene was assessed using the Four-Point Probe (Ossila). Contact angle and surface tension measurements were conducted using Theta Optical Tensiometers (Biolin Scientific) in Sessile drop mode. The thickness of spray-coated QDs was measured using a DektakXT® stylus profilometer.

**References**

[1] D. Ha, Z. Fang, and N. B. Zhitenev, "Paper in Electronic and Optoelectronic Devices," *Adv Electron Mater*, vol. 4, no. 5, p. 1700593, May 2018, doi: 10.1002/AELM.201700593.

[2] D. Tobjörk and R. Österbacka, "Paper Electronics," *Advanced Materials*, vol. 23, no. 17, pp. 1935–1961, May 2011, doi: 10.1002/ADMA.201004692.

[3] H. Zhu *et al.*, "A gravure printed antenna on shape-stable transparent nanopaper," *Nanoscale*, vol. 6, no. 15, pp. 9110–9115, Jul. 2014, doi: 10.1039/C4NR02036G.

[4] L. Yuan, B. Yao, B. Hu, K. Huo, W. Chen, and J. Zhou, "Polypyrrole-coated paper for flexible solid-state energy storage," *Energy Environ Sci*, vol. 6, no. 2, pp. 470–476, Jul. 2013, doi: 10.1039/C2EE23977A.

[5] S. Conti *et al.*, "Low-voltage 2D materials-based printed field-effect transistors for integrated digital and analog electronics on paper," *Nature Communications 2020 11:1*, vol. 11, no. 1, pp. 1–9, Jul. 2020, doi: 10.1038/s41467-020-17297-z.

[6] M. Hassan *et al.*, "Significance of Flexible Substrates for Wearable and Implantable Devices: Recent Advances and Perspectives," *Adv Mater Technol*, vol. 7, no. 3, p. 2100773, Mar. 2022, doi: 10.1002/ADMT.202100773.

[7] E. Ozden-Yenigun, I. I. Labiano, M. S. Ergoktas, A. Toomey, C. Kocabas, and A. Alomainy, "Soft Graphene-Based Antennas for Ultrawideband Wireless Communication," *ITMC Conference and Smart Textiles Salon*. Marrakesh , 2019.

[8] T. Araki *et al.*, "Broadband Photodetectors and Imagers in Stretchable Electronics Packaging," *Advanced Materials*, p. 2304048, 2023, doi: 10.1002/ADMA.202304048.

[9] Z. Duan *et al.*, "Facile, Flexible, Cost-Saving, and Environment-Friendly Paper-Based Humidity Sensor for Multifunctional Applications," *ACS Appl Mater Interfaces*, vol. 11, no. 24, pp. 21840–21849, Jun. 2019, doi: 10.1021/ACSAMI.9B05709/SUPPL_FILE/AM9B05709_SI_004.AVI.

[10] N. Sakhuja, R. Kumar, P. Katare, and N. Bhat, "Structure-Driven, Flexible, Multilayered, Paper-Based Pressure Sensor for Human-Machine Interfacing," *ACS Sustain Chem Eng*, vol. 10, no. 30, pp. 9697–9706, Aug. 2022, doi: 10.1021/ACSSUSCHEMENG.1C08491/SUPPL_FILE/SC1C08491_SI_003.AVI.

[11] Y. Zhang *et al.*, "Flexible Electronics Based on Micro/Nanostructured Paper," *Advanced Materials*, vol. 30, no. 51, p. 1801588, Dec. 2018, doi: 10.1002/ADMA.201801588.

[12] M. K. Akbari, S. Zhuiykov, S. U. Zschieschang, H. Klauk, and U. Zschieschang, "Organic transistors on paper: a brief review," *J Mater Chem C Mater*, vol. 7, no. 19, pp. 5522–5533, May 2019, doi: 10.1039/C9TC00793H.

[13]  T. Leng *et al.*, "Printed graphene/WS2 battery-free wireless photosensor on papers," *2d Mater*, vol. 7, no. 2, p. 024004, Jan. 2020, doi: 10.1088/2053-1583/AB602F.

[14]  M. Pudas, N. Halonen, P. Granat, and J. Vähäkangas, "Gravure printing of conductive particulate polymer inks on flexible substrates," *Prog Org Coat*, vol. 54, no. 4, pp. 310–316, Dec. 2005, doi: 10.1016/J.PORGCOAT.2005.07.008.

[15]  A. Vena *et al.*, "Design of chipless RFID tags printed on paper by flexography," *IEEE Trans Antennas Propag*, vol. 61, no. 12, pp. 5868–5877, 2013, doi: 10.1109/TAP.2013.2281742.

[16]  Y. Dong *et al.*, "An all-inkjet-printed flexible UV photodetector," *Nanoscale*, vol. 9, no. 25, pp. 8580–8585, Jun. 2017, doi: 10.1039/C7NR00250E.

[17]  V. T. Tran, Y. Wei, H. Yang, Z. Zhan, and H. Du, "All-inkjet-printed flexible ZnO micro photodetector for a wearable UV monitoring device," *Nanotechnology*, vol. 28, no. 9, p. 095204, Jan. 2017, doi: 10.1088/1361-6528/AA57AE.

[18]  G. Hu *et al.*, "Functional inks and printing of two-dimensional materials," *Chem Soc Rev*, vol. 47, no. 9, pp. 3265–3300, May 2018, doi: 10.1039/C8CS00084K.

[19]  S. Pinilla, J. Coelho, K. Li, J. Liu, and V. Nicolosi, "Two-dimensional material inks," *Nature Reviews Materials 2022 7:9*, vol. 7, no. 9, pp. 717–735, May 2022, doi: 10.1038/s41578-022-00448-7.

[20]  S. Conti *et al.*, "Low-voltage 2D materials-based printed field-effect transistors for integrated digital and analog electronics on paper," *Nature Communications 2020 11:1*, vol. 11, no. 1, pp. 1–9, Jul. 2020, doi: 10.1038/s41467-020-17297-z.

[21]  Y. Hernandez *et al.*, "High-yield production of graphene by liquid-phase exfoliation of graphite," *Nature Nanotechnology 2008 3:9*, vol. 3, no. 9, pp. 563–568, Aug. 2008, doi: 10.1038/nnano.2008.215.

[22]  M. Jafarpour, F. Nüesch, J. Heier, and S. Abdolhosseinzadeh, "Functional Ink Formulation for Printing and Coating of Graphene and Other 2D Materials: Challenges and Solutions," *Small Science*, vol. 2, no. 11, p. 2200040, Nov. 2022, doi: 10.1002/SMSC.202200040.

[23]  S. Akhavan *et al.*, "Graphene-black phosphorus printed photodetectors," *2d Mater*, vol. 10, no. 3, p. 035015, May 2023, doi: 10.1088/2053-1583/ACC74C.

[24]  K. Cho, T. Lee, and S. Chung, "Inkjet printing of two-dimensional van der Waals materials: a new route towards emerging electronic device applications," *Nanoscale Horiz*, vol. 7, no. 10, pp. 1161–1176, Sep. 2022, doi: 10.1039/D2NH00162D.

[25]  X. Feng *et al.*, "Spray-coated perovskite hemispherical photodetector featuring narrow-band and wide-angle imaging," *Nature Communications 2022 13:1*, vol. 13, no. 1, pp. 1–9, Oct. 2022, doi: 10.1038/s41467-022-33934-1.

[26]  A. Sarycheva, A. Polemi, Y. Liu, K. Dandekar, B. Anasori, and Y. Gogotsi, "2D titanium carbide (MXene) for wireless communication," *Sci Adv*, vol. 4, no. 9, Sep. 2018, doi: 10.1126/SCIADV.AAU0920/SUPPL_FILE/AAU0920_SM.PDF.

[27]  K. Lobo, R. Thakur, S. K. Prasad, and H. S. S. R. Matte, "Solution-processed 2D materials on paper substrates for photodetection and photomechanical applications," *J Mater Chem C Mater*, vol. 10, no. 48, pp. 18326–18335, Dec. 2022, doi: 10.1039/D2TC02742A.

[28]  C. J. Zhang *et al.*, "Transparent, Flexible, and Conductive 2D Titanium Carbide (MXene) Films with High Volumetric Capacitance," *Advanced Materials*, vol. 29, no. 36, p. 1702678, Sep. 2017, doi: 10.1002/ADMA.201702678.

[29] W. Qian *et al.*, "An aerosol-liquid-solid process for the general synthesis of halide perovskite thick films for direct-conversion X-ray detectors," *Matter*, vol. 4, no. 3, pp. 942–954, Mar. 2021, doi: 10.1016/J.MATT.2021.01.020.

[30] Y. Zhang *et al.*, "Flexible Electronics Based on Micro/Nanostructured Paper," *Advanced Materials*, vol. 30, no. 51, p. 1801588, Dec. 2018, doi: 10.1002/ADMA.201801588.

[31] W. Deng *et al.*, "All-Sprayed-Processable, Large-Area, and Flexible Perovskite/MXene-Based Photodetector Arrays for Photocommunication," *Adv Opt Mater*, vol. 7, no. 6, p. 1801521, Mar. 2019, doi: 10.1002/ADOM.201801521.

[32] S. X. Li, X. L. Xu, Y. Yang, Y. S. Xu, Y. Xu, and H. Xia, "Highly Deformable High-Performance Paper-Based Perovskite Photodetector with Improved Stability," *ACS Appl Mater Interfaces*, vol. 13, no. 27, pp. 31919–31927, Jul. 2021, doi: 10.1021/ACSAMI.1C05828/ASSET/IMAGES/LARGE/AM1C05828_0006.JPEG.

[33] S. Akhavan *et al.*, "Multiexciton generation assisted highly photosensitive CdHgTe nanocrystal skins," *Nano Energy*, vol. 26, pp. 324–331, Aug. 2016, doi: 10.1016/J.NANOEN.2016.04.055.

[34] S. Akhavan *et al.*, "Flexible and fragmentable tandem photosensitive nanocrystal skins," *Nanoscale*, vol. 8, no. 8, pp. 4495–4503, Feb. 2016, doi: 10.1039/C5NR05063D.

[35] S. Akhavan, A. Yeltik, and H. V. Demir, "Photosensitivity enhancement with TiO2 in semitransparent light-sensitive skins of nanocrystal monolayers," *ACS Appl Mater Interfaces*, vol. 6, no. 12, pp. 9023–9028, Jun. 2014, doi: 10.1021/AM502472Y/SUPPL_FILE/AM502472Y_SI_001.PDF.

[36] G. Konstantatos *et al.*, "Hybrid graphene–quantum dot phototransistors with ultrahigh gain," *Nature Nanotechnology 2012 7:6*, vol. 7, no. 6, pp. 363–368, May 2012, doi: 10.1038/nnano.2012.60.

[37] Y. Lee *et al.*, "High-Performance Perovskite–Graphene Hybrid Photodetector," *Advanced Materials*, vol. 27, no. 1, pp. 41–46, Jan. 2015, doi: 10.1002/ADMA.201402271.

[38] F. Brunetti *et al.*, "Printed Solar Cells and Energy Storage Devices on Paper Substrates," *Adv Funct Mater*, vol. 29, no. 21, p. 1806798, May 2019, doi: 10.1002/ADFM.201806798.

[39] J. Lessing, A. C. Glavan, S. B. Walker, C. Keplinger, J. A. Lewis, and G. M. Whitesides, "Inkjet Printing of Conductive Inks with High Lateral Resolution on Omniphobic 'RF Paper' for Paper-Based Electronics and MEMS," *Advanced Materials*, vol. 26, no. 27, pp. 4677–4682, Jul. 2014, doi: 10.1002/ADMA.201401053.

[40] Q. A. Akkerman *et al.*, "Strongly emissive perovskite nanocrystal inks for high-voltage solar cells," *Nature Energy 2016 2:2*, vol. 2, no. 2, pp. 1–7, Dec. 2016, doi: 10.1038/nenergy.2016.194.

[41] L. Protesescu *et al.*, "Nanocrystals of Cesium Lead Halide Perovskites (CsPbX3, X = Cl, Br, and I): Novel Optoelectronic Materials Showing Bright Emission with Wide Color Gamut," *Nano Lett*, vol. 15, no. 6, pp. 3692–3696, Jun. 2015, doi: 10.1021/NL5048779/SUPPL_FILE/NL5048779_SI_001.PDF.

[42] Q. A. Akkerman *et al.*, "Tuning the optical properties of cesium lead halide perovskite nanocrystals by anion exchange reactions," *J Am Chem Soc*, vol. 137, no. 32, pp. 10276–10281, Aug. 2015, doi: 10.1021/JACS.5B05602/SUPPL_FILE/JA5B05602_SI_003.AVI.

[43] L. Dou *et al.*, "Atomically thin two-dimensional Organic-inorganic hybrid perovskites," *Science (1979)*, vol. 349, no. 6255, pp. 1518–1521, Sep. 2015, doi: 10.1126/SCIENCE.AAC7660/SUPPL_FILE/AAC7660-DOU.SM.PDF.

[44] V. D'Innocenzo, A. R. Srimath Kandada, M. De Bastiani, M. Gandini, and A. Petrozza, "Tuning the light emission properties by band gap engineering in hybrid lead halide perovskite," *J Am Chem Soc*, vol. 136, no. 51, pp. 17730–17733, Dec. 2014, doi: 10.1021/JA511198F/SUPPL_FILE/JA511198F_SI_001.PDF.

[45] Z. Zhu *et al.*, "Efficiency enhancement of perovskite solar cells through fast electron extraction: The role of graphene quantum dots," *J Am Chem Soc*, vol. 136, no. 10, pp. 3760–3763, Mar. 2014, doi: 10.1021/JA4132246/SUPPL_FILE/JA4132246_SI_001.PDF.

[46] H. C. Woo, J. W. Choi, J. Shin, S. H. Chin, M. H. Ann, and C. L. Lee, "Temperature-Dependent Photoluminescence of CH3NH3PbBr3 Perovskite Quantum Dots and Bulk Counterparts," *J Phys Chem Lett*, vol. 9, no. 14, pp. 4066–4074, Jul. 2018, doi: 10.1021/ACS.JPCLETT.8B01593.

[47] J. S. Chen, T. L. Doane, M. Li, H. Zang, M. M. Maye, and M. Cotlet, "0D–2D and 1D–2D Semiconductor Hybrids Composed of All Inorganic Perovskite Nanocrystals and Single-Layer Graphene with Improved Light Harvesting," *Particle & Particle Systems Characterization*, vol. 35, no. 2, p. 1700310, Feb. 2018, doi: 10.1002/PPSC.201700310.

[48] Beer, "Bestimmung der Absorption des rothen Lichts in farbigen Flüssigkeiten," *Ann Phys*, vol. 162, no. 5, pp. 78–88, Jan. 1852, doi: 10.1002/ANDP.18521620505.

[49] Q. Li, J. S. Church, A. Kafi, M. Naebe, and B. L. Fox, "An improved understanding of the dispersion of multi-walled carbon nanotubes in non-aqueous solvents," *Journal of Nanoparticle Research*, vol. 16, no. 7, pp. 1–12, Jun. 2014, doi: 10.1007/S11051-014-2513-0/METRICS.

[50] F. Bonaccorso, P. H. Tan, and A. C. Ferrari, "Multiwall nanotubes, multilayers, and hybrid nanostructures: New frontiers for technology and Raman spectroscopy," *ACS Nano*, vol. 7, no. 3, pp. 1838–1844, Mar. 2013, doi: 10.1021/NN400758R/ASSET/IMAGES/MEDIUM/NN-2013-00758R_0005.GIF.

[51] G. E. J. Poinern, D. Parsonage, T. B. Issa, M. K. Ghosh, E. Paling, and P. Singh, "Preparation, characterization and As(V) adsorption behaviour of CNT-ferrihydrite composites," *International Journal of Engineering, Science and Technology*, vol. 2, no. 8, pp. 13–24, Feb. 2010, doi: 10.4314/IJEST.V2I8.63776.

[52] V. R. Nazeera Banu, V. Ramesh Babu, and S. Rajendran, "INVESTIGATING THE CORROSION INHIBITION EFFICIENCY OF SURGICAL CARBON STEEL INSTRUMENTS USED IN MEDICAL FIELD," *Int. Res. J. Pharm*, vol. 2017, no. 12, doi: 10.7897/2230-8407.0812254.

[53] A. A. Lagatsky *et al.*, "2 μm solid-state laser mode-locked by single-layer graphene," *Appl Phys Lett*, vol. 102, no. 1, Jan. 2013, doi: 10.1063/1.4773990/126114.

[54] A. C. Ferrari *et al.*, "Raman spectrum of graphene and graphene layers," *Phys Rev Lett*, vol. 97, no. 18, p. 187401, Oct. 2006, doi: 10.1103/PHYSREVLETT.97.187401/FIGURES/3/MEDIUM.

[55] A. C. Ferrari and D. M. Basko, "Raman spectroscopy as a versatile tool for studying the properties of graphene," *Nat Nanotechnol*, vol. 8, no. 4, pp. 235–246, 2013, doi: 10.1038/NNANO.2013.46.

[56] M. U. Memon, M. M. Tentzeris, and S. Lim, "Inkjet-printed 3D Hilbert-curve fractal antennas for VHF band," *Microw Opt Technol Lett*, vol. 59, no. 7, pp. 1698–1704, Jul. 2017, doi: 10.1002/MOP.30613.

[57] L. G. Cançado *et al.*, "Quantifying defects in graphene via Raman spectroscopy at different excitation energies," *Nano Lett*, vol. 11, no. 8, pp. 3190–3196, Aug. 2011, doi: 10.1021/NL201432G/ASSET/IMAGES/MEDIUM/NL-2011-01432G_0007.GIF.

[58] M. Bruna, A. K. Ott, M. Ijäs, D. Yoon, U. Sassi, and A. C. Ferrari, "Doping dependence of the Raman spectrum of defected graphene," *ACS Nano*, vol. 8, no. 7, pp. 7432–7441, Jul. 2014, doi: 10.1021/NN502676G/ASSET/IMAGES/MEDIUM/NN-2014-02676G_0011.GIF.

[59] S. Lee, K. Lee, C.-H. Liu, and Z. Zhong, "Homogeneous bilayer graphene film based flexible transparent conductor".

[60] S. M. Sze and K. K. Ng, "Physics of Semiconductor Devices," *Physics of Semiconductor Devices*, Oct. 2006, doi: 10.1002/0470068329.

[61] S. M. Sze and K. K. Ng, "Physics of Semiconductor Devices," *Physics of Semiconductor Devices*, Oct. 2006, doi: 10.1002/0470068329.

[62] G. Konstantatos *et al.*, "Hybrid graphene–quantum dot phototransistors with ultrahigh gain," *Nature Nanotechnology 2012 7:6*, vol. 7, no. 6, pp. 363–368, May 2012, doi: 10.1038/nnano.2012.60.

[63] Y. Liu *et al.*, "Highly Efficient and Air-Stable Infrared Photodetector Based on 2D Layered Graphene-Black Phosphorus Heterostructure," *ACS Appl Mater Interfaces*, vol. 9, no. 41, pp. 36137–36145, Oct. 2017, doi: 10.1021/ACSAMI.7B09889/SUPPL_FILE/AM7B09889_SI_001.PDF.

[64] K. Arora, K. Kaur, and M. Kumar, "Superflexible, Self-Biased, High-Voltage-Stable, and Seal-Packed Office-Paper Based Gallium-Oxide Photodetector," *ACS Appl Electron Mater*, vol. 3, no. 4, pp. 1852–1863, Apr. 2021, doi: 10.1021/ACSAELM.1C00099/ASSET/IMAGES/LARGE/EL1C00099_0006.JPEG.

[65] S. Veeralingam and S. Badhulika, "Paper-based flexible, VIS-NIR photodetector with actively variable spectrum and enhanced responsivity using surface engineered transitional metal buffer layer," *FlatChem*, vol. 33, p. 100370, May 2022, doi: 10.1016/J.FLATC.2022.100370.

[66] S. Cai, C. Zuo, J. Zhang, H. Liu, and X. Fang, "A Paper-Based Wearable Photodetector for Simultaneous UV Intensity and Dosage Measurement," *Adv Funct Mater*, vol. 31, no. 20, May 2021, doi: 10.1002/ADFM.202100026.

[67] A. Mazaheri, M. Lee, H. S. J. Van Der Zant, R. Frisenda, and A. Castellanos-Gomez, "MoS2-on-paper optoelectronics: drawing photodetectors with van der Waals semiconductors beyond graphite," *Nanoscale*, vol. 12, no. 37, pp. 19068–19074, Oct. 2020, doi: 10.1039/D0NR02268C.

[68] N. J. A. Cordeiro *et al.*, "Fast and Low-Cost Synthesis of MoS2 Nanostructures on Paper Substrates for Near-Infrared Photodetectors," *Applied Sciences 2021, Vol. 11, Page 1234*, vol. 11, no. 3, p. 1234, Jan. 2021, doi: 10.3390/APP11031234.

[69] P. T. Gomathi, P. Sahatiya, and S. Badhulika, "Large-Area, Flexible Broadband Photodetector Based on ZnS–MoS2 Hybrid on Paper Substrate," *Adv Funct Mater*, vol. 27, no. 31, Aug. 2017, doi: 10.1002/ADFM.201701611.

[70] C. H. Lin *et al.*, "Highly Deformable Origami Paper Photodetector Arrays," *ACS Nano*, vol. 11, no. 10, pp. 10230–10235, Oct. 2017, doi: 10.1021/ACSNANO.7B04804.

[71] H. Li, S. Jiao, H. Li, L. Li, and X. Zhang, "Band-edge modulated ZnO pomegranates-on-paper photodetector," *J Mater Chem C Mater*, vol. 3, no. 15, pp. 3702–3707, Apr. 2015, doi: 10.1039/C4TC02787F.

[72] H. Fang et al., "An Origami Perovskite Photodetector with Spatial Recognition Ability," *ACS Appl Mater Interfaces*, vol. 9, no. 12, pp. 10921–10928, Mar. 2017, doi: 10.1021/ACSAMI.7B02213/ASSET/IMAGES/LARGE/AM-2017-02213Q_0004.JPEG.

[73] P. M. Pataniya and C. K. Sumesh, "WS2Nanosheet/Graphene Heterostructures for Paper-Based Flexible Photodetectors," *ACS Appl Nano Mater*, vol. 3, no. 7, pp. 6935–6944, Jul. 2020, doi: 10.1021/ACSANM.0C01276.

[74] P. Pataniya et al., "Paper-Based Flexible Photodetector Functionalized by WSe2 Nanodots," *ACS Appl Nano Mater*, vol. 2, no. 5, pp. 2758–2766, May 2019, doi: 10.1021/ACSANM.9B00266.

[75] V. Selamneni et al., "MoS2/Paper Decorated with Metal Nanoparticles (Au, Pt, and Pd) Based Plasmonic-Enhanced Broadband (Visible-NIR) Flexible Photodetectors," *Adv Mater Interfaces*, vol. 8, no. 6, p. 2001988, Mar. 2021, doi: 10.1002/ADMI.202001988.

[76] P. M. Pataniya, V. Patel, and C. K. Sumesh, "MoS2/WSe2 nanohybrids for flexible paper-based photodetectors," *Nanotechnology*, vol. 32, no. 31, p. 315709, May 2021, doi: 10.1088/1361-6528/ABF77A.

[77] S. Veeralingam and S. Badhulika, "Bi-Metallic sulphides 1D Bi2S3 microneedles/1D RuS2 nano-rods based n-n heterojunction for large area, flexible and high-performance broadband photodetector," *J Alloys Compd*, vol. 885, p. 160954, Dec. 2021, doi: 10.1016/J.JALLCOM.2021.160954.

[78] T. Zhang et al., "Six-arm Stellat Dendritic-PbS Flexible Infrared Photodetector for Intelligent Healthcare Monitoring," *Adv Mater Technol*, vol. 7, no. 8, p. 2200250, Aug. 2022, doi: 10.1002/ADMT.202200250.

[79] A. Verma, P. Chaudhary, A. Singh, R. K. Tripathi, and B. C. Yadav, "ZnS Nanosheets in a Polyaniline Matrix as Metallopolymer Nanohybrids for Flexible and Biofriendly Photodetectors," *ACS Appl Nano Mater*, vol. 5, no. 4, pp. 4860–4874, Apr. 2022, doi: 10.1021/ACSANM.1C04437/ASSET/IMAGES/LARGE/AN1C04437_0011.JPEG.

[80] D. Xiong et al., "Controllable in-situ-oxidization of 3D-networked Ti3C2Tx-TiO2 photodetectors for large-area flexible optical imaging," *Nano Energy*, vol. 93, p. 106889, Mar. 2022, doi: 10.1016/J.NANOEN.2021.106889.

[81] Y. Sun, Z. Liu, and Y. Ding, "Paper-based flexible broadband photodetectors functionalized by PbS quantum dots/Carbon nanotube networks hybrid structure," IEEE International Conference on Integrated Circuits, Technologies and Applications. Accessed: Aug. 29, 2023. [Online]. Available: https://ieeexplore.ieee.org/stamp/stamp.jsp?arnumber=9661965

[82] D. Kannichankandy et al., "Paper based organic–inorganic hybrid photodetector for visible light detection," *Appl Surf Sci*, vol. 524, p. 146589, Sep. 2020, doi: 10.1016/J.APSUSC.2020.146589.

[83] V. Selamneni, T. Akshaya, V. Adepu, and P. Sahatiya, "Laser-assisted micropyramid patterned PDMS encapsulation of 1D tellurium nanowires on cellulose paper for highly sensitive strain sensor and its photodetection studies," *Nanotechnology*, vol. 32, no. 45, p. 455201, Aug. 2021, doi: 10.1088/1361-6528/AC19D8.

[84] V. Selamneni, S. Kanungo, and P. Sahatiya, "Large area growth of SnS 2 /graphene on cellulose paper as a flexible broadband photodetector and investigating its band structure through first principles calculations," *Mater Adv*, vol. 2, no. 7, pp. 2373–2381, Apr. 2021, doi: 10.1039/D1MA00054C.

[85] Z. Yang *et al.*, "Graphene Textile Strain Sensor with Negative Resistance Variation for Human Motion Detection," *ACS Nano*, vol. 12, no. 9, pp. 9134–9141, Sep. 2018, doi: 10.1021/ACSNANO.8B03391/SUPPL_FILE/NN8B03391_SI_001.PDF.

[86] F. Pan *et al.*, "3D Graphene Films Enable Simultaneously High Sensitivity and Large Stretchability for Strain Sensors," *Adv Funct Mater*, vol. 28, no. 40, p. 1803221, Oct. 2018, doi: 10.1002/ADFM.201803221.

[87] P. Di Qi *et al.*, "Understanding the Cycling Performance Degradation Mechanism of a Graphene-Based Strain Sensor and an Effective Corresponding Improvement Solution," *ACS Appl Mater Interfaces*, vol. 12, no. 20, pp. 23272–23283, May 2020, doi: 10.1021/ACSAMI.0C00176/SUPPL_FILE/AM0C00176_SI_007.AVI.

# Supplementary Information

**Spray-Coated Graphene/Quantum Dots Paper-Based Photodetectors**

S. Malik[1], Y. Zhao[1], Y. He[1], X. Zhao[1], H. Li[1], L. G. Occhipinti[2], M. Wang[1], and S. Akhavan[1*]

[1]Institute for Materials Discovery, University College London, Torrington Place, London, WC1E 7JE, UK

[2]Cambridge Graphene Centre, University of Cambridge, JJ Thompson Avenue, Cambridge CB3 0FA, UK

Supplementary Information 1

Quantum dots (QDs) were spray coated on paper substrate aiming for an approximate thickness of 500 nm.

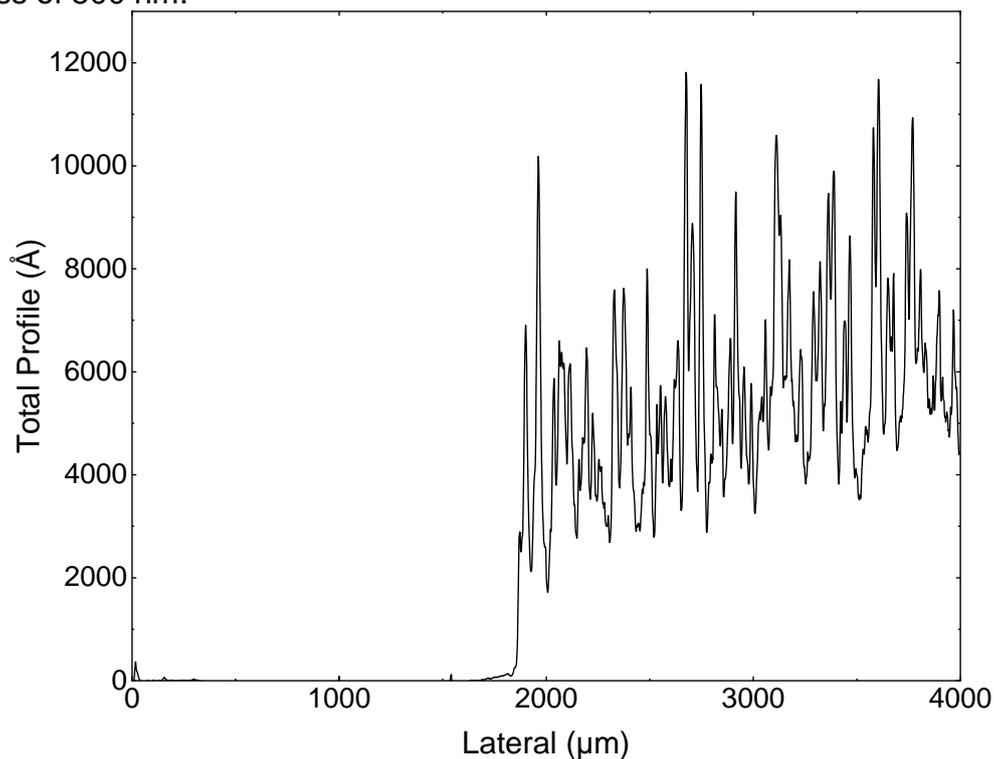

Fig. S1. The thickness profile of spray coated QDs with a target thickness of approximately 500 nm.

Supplementary Information 2

Optical contact angle measurements (Attension) were employed for characterization. The contact angle measurement demonstrated a 25.49⁰ value, with a surface tension of 235 mN/m observed for a PS drop on SLG that was transferred onto the paper substrate, Fig. S2a. The PS solution's wettability showcases its function as a protective coating, effectively adhering to the SLG and safeguarding the CVD SLG during plasma etching.

To evaluate the suitability of the electrode, we examined the wettability of CNT ink to transferred SLG on a paper substrate. We used optical contact angle and surface tension measurements (Attension) for analysis. The contact angle measured 29.51⁰ and the surface tension was 224.3 mN/m for the CNT drop on SLG transferred onto the paper substrate. This facilitated rapid drying with limited spreading beyond the specified area, Fig. S2b.

Additionally, a contact angle of 10.88° and a surface tension of 329.3 mN/m were recorded for a QDs drop on SLG, Fig. S2c. This minimal contact angle indicates a robust attraction between the SLG surface and QDs.

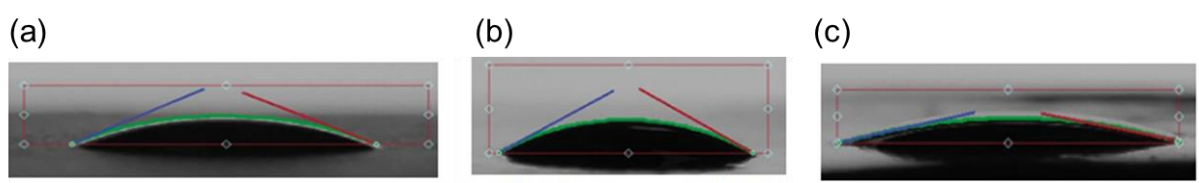

Fig. S2. Shape of the droplet and contact angle measurements of (a) PS ink, (b) CNT ink, and (c) $CsPbBr_3$ perovskite QDs.

Supplementary Information 3

The scanning electron microscope image illustrates the extent of dispersion and clustering of the CNTs applied to the paper substrate.

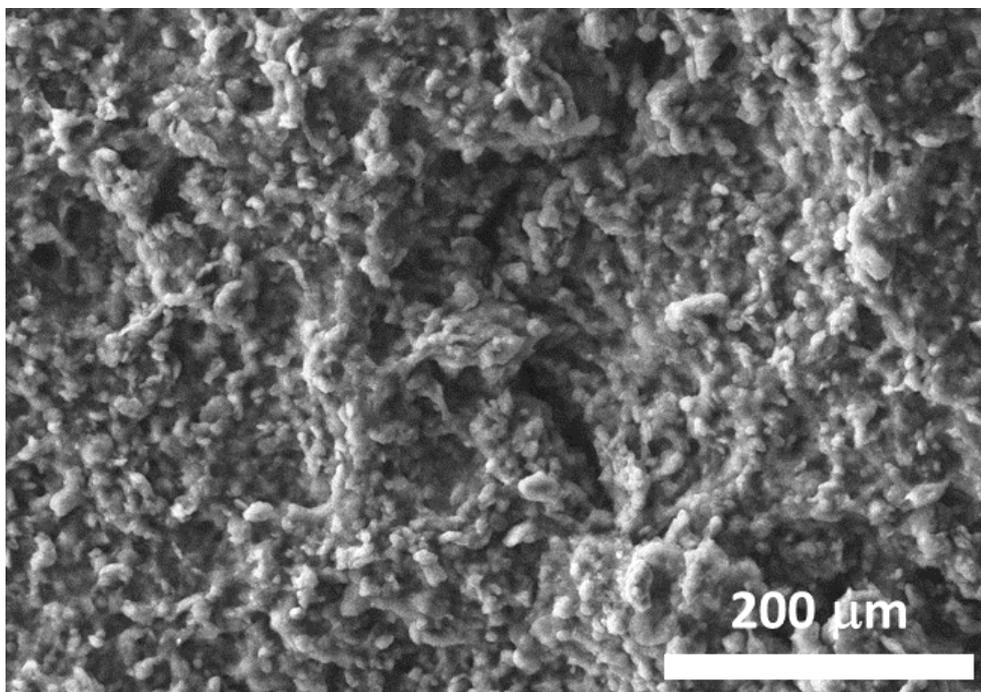

Fig. S3. Scanning electron microscopy image of spray-coated CNTs on paper substrate.